\documentclass[journal=jpccck,manuscript=letter]{achemso}
\setkeys{acs}{email=false}
\usepackage[english]{babel}
\usepackage[T1]{fontenc}
\usepackage{csquotes}
\usepackage[fleqn]{amsmath}
\usepackage{amsmath}
\usepackage{setspace}
\usepackage{pdfpages}
\usepackage{wasysym}
\usepackage{amsmath}
\usepackage{mhchem}
\usepackage{graphicx}
\usepackage{subcaption}
\usepackage{textcomp}

\usepackage{array}
\title{}
\newcommand{\vv}[1]{}

\title{Benchmarking conductivity predictions of the Advanced Electrolyte Model (AEM) for aqueous systems}
\author{Adarsh Dave}
\affiliation{Department of Mechanical Engineering, Carnegie Mellon University, Pittsburgh, Pennsylvania, 15213, USA.\\
}
\altaffiliation{Wilton E. Scott Institute for Energy Innovation, Carnegie Mellon University, Pittsburgh, Pennsylvania, 15213, USA.}

\author{Kevin L. Gering}
\affiliation{Idaho National Laboratory, 2525 Fremont Ave, Idaho Falls, ID 83402.\\
}

\author{Jared M. Mitchell}
\affiliation{Department of Materials Science and Engineering, Carnegie Mellon University, Pittsburgh, Pennsylvania, 15213, USA.\\
}
\altaffiliation{Wilton E. Scott Institute for Energy Innovation, Carnegie Mellon University, Pittsburgh, Pennsylvania, 15213, USA.}

\author{Jay Whitacre}
\affiliation{Department of Materials Science and Engineering, Carnegie Mellon University, Pittsburgh, Pennsylvania, 15213, USA.\\
}
\altaffiliation{Wilton E. Scott Institute for Energy Innovation, Carnegie Mellon University, Pittsburgh, Pennsylvania, 15213, USA.}

\author{Venkatasubramanian Viswanathan}
\affiliation{Department of Mechanical Engineering, Carnegie Mellon University, Pittsburgh, Pennsylvania, 15213, USA.\\
}
\altaffiliation{Wilton E. Scott Institute for Energy Innovation, Carnegie Mellon University, Pittsburgh, Pennsylvania, 15213, USA.}
\email{venkvis@cmu.edu}

\begin{document}

\onehalfspacing
\maketitle
\begin{abstract}

High-concentration aqueous electrolytes have shown promise as candidates for a safer, lower-cost battery system. Ionic conductivity is a key property required in high performing electrolytes; the Advanced Electrolyte Model (AEM) has previously shown great accuracy in predicting ionic conductivity in highly-concentrated non-aqueous electrolytes. This work provides extensive experimental data for mixed and highly concentrated aqueous electrolyte systems, rapidly generated via a robotic electrolyte testing apparatus. These data demonstrate exceptional accuracy from AEM in predicting conductivity in aqueous systems, with the accuracy being maintained even in highly-concentrated and mixed-salt regimes.
\end{abstract}

\section{Introduction}

Batteries are playing a crucial role in driving the transition to electrification of passenger vehicles. However, there is a need to improve current Li-ion batteries to meet the needs for electric pickup and semi-trucks, electric vertical take-off and landing aircraft (eVTOL), long-duration energy storage for the grid, etc. \cite{Sripad2017,Sripad2017a,Li2017,Guttenberg2017,Fredericks2018}. Organic electrolytes currently used in Li-ion batteries limit the cycle life, safety, energy density, etc. \cite{Arbizzani2011,Roth2016,Khetan2014,Khetan2015}. Aqueous electrolytes offer significantly improved safety but have typically been limited to a small voltage window, limiting the energy density and cycle life for aqueous batteries ($\sim$ 2V) \cite{Suo2016}.

Recent works have shown that aqueous electrolytes with high salt concentration can lead to vastly improved voltage stability \cite{Suo2016,Suo2016a,Suo2017,Suo2017a,Yang2017,Lukatskaya2018}.  This offers great promise for optimization of aqueous electrolytes for many real-world applications \cite{Wessells2010,Whitacre2012,Wu2015,Wang2016,Wang2017,Yang2019}. In addition to voltage stability, transport properties are an important criterion determining power capability of batteries \cite{Plichta2000,Lukatskaya2018}. As an example, many non-aqueous Li-ion electrolytes have RT conductivity of around 10 mS/cm, which is considered acceptable.  Yet, many aqueous electrolytes offer conductivities beyond 100 mS/cm with excellent salt solubility past 5 molal. Given the enormous space of possible concentrated electrolytes, it is important to have a screening tool that can accurately predict the transport properties for complex aqueous electrolytes.

The Advanced Electrolyte Model (AEM) provides this capability using a rigorous statistical thermodynamics approach and has been proved to be a very effective tool for optimizing organic electrolytes \cite{Gering2017,Logan2018}.  A similarly thorough evaluation of AEM for accuracy predicting aqueous electrolyte properties has not been demonstrated before.  In this work, we demonstrate the reliability of AEM to predict conductivity of concentrated electrolyte solutions consisting of unary and binary salt mixtures.  Leveraging a highly-automated robotic test stand, we critically evaluate AEM over a range of relevant concentrated electrolytes. We find exceptional accuracy from AEM in predicting conductivity in aqueous systems, even in highly-concentrated and mixed-salt systems. Maximum deviation from experimental conductivity values is near 10\%, which is similar to previous conductivity benchmarks of AEM in concentrated non-aqueous systems \cite{Gering2017}.  These results indicate that the AEM can be used as a valuable tool in optimizing aqueous electrolyte formulations and as a useful prior for optimal design of experiments.

\section{Methods}
\subsection{Advanced Electrolyte Model}

AEM presents a rigorous framework for predicting macroscopic properties of electrolytes based on molecular interactions and quantities under realistic system conditions (i.e. highly mixed solvents and salts at varying temperatures). Although the formalism is universal across aqueous and non-aqueous systems, AEM's conductivity predictions have been primarily validated for non-aqueous systems \cite{Gering2017}, leaving aqueous systems as a fair target for more extensive validation.  A large suite of properties is generated with each AEM simulation, ranging from transport properties, thermodynamic terms (e.g., ion solvation energies), sizes of solvated species, ion association, physical properties, permittivity metrics, preferential ion solvation, terms for Walden analysis, and others.

AEM models electrolyte species as a function of temperature and concentration via the non-primitive non-restricted associative form of the Mean Spherical Approximation (NPNRAMSA)\cite{Gering2006}. Thermodynamic governing equations derived from the Mass Action Law (MAL) are parametrized with ion size and solvation quantities, solution densities, permittivity, and ionic number densities and solved as a function of composition, concentration, and temperature, yielding estimates of ion species present in the electrolyte (e.g. contact ion-pairs, solvent-separated ion-pairs)\cite{Gering2006}. The central solvation parameters include solvent residence time and average ion-solvent ligand distance - these are obtained from quantum chemistry simulations, experimental measurements, or other forms of modeling\cite{Gering2006}. The `non-primitive' portion refers to treatment of permittivity as colligative property, with changes in permittivity with concentration explicitly derived in relation to MAL\cite{Gering2017}.

AEM models conductivity by correcting the Stokes-Einstein law for conditions present in highly non-ideal electrolyte systems, analogous to correcting the ideal gas law in thermodynamics for interacting systems\cite{Thorn2007}. The central solvation quantities are used to estimate the effective solvated ionic radii, used as the transport size of the ion in the corrected Stokes equation. Amendments to the Stokes equation were derived specifically for this model to include 7 different physical contributions important to transport in highly-concentrated electrolytes, including: ion-solvent interactions via solvation energetics and solvated ion sizes; ion-ion interactions via ion association and electro-static models; counter-ion transport; ionic hopping; ionic random motion; and viscosity as a function of composition and concentration. Additional information about specific physical models capturing each of these contributions can be found elsewhere\cite{Gering2017}.

AEM models viscosity by accounting for structure-making, structure-breaking, ion-ion interaction, and solvent-ion interaction contributions to changes in an electrolyte viscosity. Starting first with the pure solvent viscosity, ion species populations are estimated with NPNRMSA; specific species fractions are used in models for solvent structure-making and structure-breaking. Electrostatic attraction between screened ions are modeled, along with the diminished solvent movement given ion fields and volumes. These factors are multiplied by the pure solvent viscosity to obtain electrolyte viscosity as a function of concentration.  For example, a greater extent of ion solvation (typically seen for cations) will result in structure-making behavior that increases viscosity.  This is demonstrated in comparing aqueous LiCl to KCl, wherein the LiCl case has a strongly solvated cation and higher viscosity while the KCl case has a lesser solvated cation and lower viscosity over salt concentration \cite{Gering2006}. Additional information about the specific physical models capturing these contributions can be found elsewhere \cite{Gering2006}.

The AEM architecture has rigorous consideration of ion association in single and mixed-salt systems.  Due to the decrease of solution permittivity over increasing salt concentration (termed "dielectric depression"), a complex iterative scheme is required to correctly match the combined extents of ion dissociation and ion association to dielectric depression.  This is all maintained on the chemical physics and thermodynamic levels, involving permittivity-reliant electrostatic functions and MAL.  The result is a robust method to estimate populations of single ions, ion pairs and triple ions at any condition of composition and temperature.  For mixed-salt electrolytes, there is further complication driven by the probability of forming prevalent types of ion pairs (i.e. given \ce{AB + AC} in solution, whether they would form \ce{AB + AC ->A+ + B- + A$^*$C} or \ce{A+ + C- + A$^*$B}).  AEM handles all mixed-salt terms on the molecular scale, and adjusts population attributes based on the outcome of the prevalent ion association behavior. For aqueous systems herein, ion association is kept low due to the high relative permittivity of water.  However, AEM has successfully described other cases involving low-permittivity solvents such as dimethyl carbonate (relative permittivity below 3 at RT), wherein solution permittivity eventually increases over salt concentration due to the contribution from ion pair dipoles to the overall permittivity \cite{Logan2018a}.

Previous work has shown a maximum of 15\% deviation between experimental values of conductivity in concentration non-aqueous electrolytes and AEM calculations \cite{Gering2017}. Viscosity calculations have yielded a less than 3\% difference for chloride salts in water \cite{Gering2006}.

\subsection{Experimental Testing}

The test-stand is mechanically automated, mixing desired electrolytes from stock solutions of single salts dissolved in water. Desired experiments are issued over a Python API to a HTTP web-server, which commands a Labview package that orchestrates the physical experiment. Measurements are taken with Consort C3410 dual-channel, 4-electrode conductivity probe. The meter compensates readings with onboard software to 20\textdegree C; lab temperature varied between 21\textdegree C and 23\textdegree C for values given in this paper. The meter was calibrated with KCl lab-standard obtained from Sigma-Aldrich, to an accuracy of 0.1 mS/cm and standard error of 0.5 mS/cm. A detailed description of the test-stand design (along with extensive studies on experimental accuracy, variance, and drift) can be found in an accompanying paper. \cite{teststand}

Lithium nitrate was obtained from Sigma-Aldrich (ReagentPlus grade), and dissolved into a stock solution of 7.02m. Sodium nitrate was obtained from Sigma-Aldrich (ACS reagent, >99.0\% grade), and dissolved into a stock solution of 10.02m. Lithium sulfate was obtained from Sigma-Aldrich (Titration >98.5\% grade), and dissolved into a stock solution of 3.01m. Sodium sulfate was obtained from Sigma-Aldrich (ACS reagent, anhydrous, redi-Dri >99.0\% grade), and dissolved into a stock solution of 1.50m. Stock solution concentrations were chosen to be near the solubility limit of the each salt in water at 20\textdegree C.

This particular salt selection is intended to give a broad sampling of salts with potential for use in high-concentration aqueous battery electrolytes \cite{Wu2015}. Lithium and sodium cation salts are ubiquitous for use in intercalation chemistry batteries, sometimes in a mixed cation battery, and generally frequently studied for high-voltage aqueous batteries \cite{Whitacre2012,Zheng2018}. Nitrate and sulfate were chosen as anions for sodium salts to contrast a high-solubility, high-conductivity salt (\ce{NaNO3}), with a low-solubility and medium-conductivity salt (\ce{Na2SO4}). The choice of anion also had different conductivity behavior with changing concentration - two were monotonically-increasing and two were not. \ce{LiNO3}/\ce{Li2SO4}, \ce{LiNO3}/\ce{NaNO3}, and \ce{LiNO3}/\ce{Na2SO4} were chosen as the binary benchmark cases to illustrate a mixed anion, mixed cation, and mixed anion/mixed cation case for AEM.
	
\section{Results and Discussion}
\begin{figure}
    \centering
    \begin{subfigure}[b]{0.4\textwidth}
        \includegraphics[width=\textwidth]{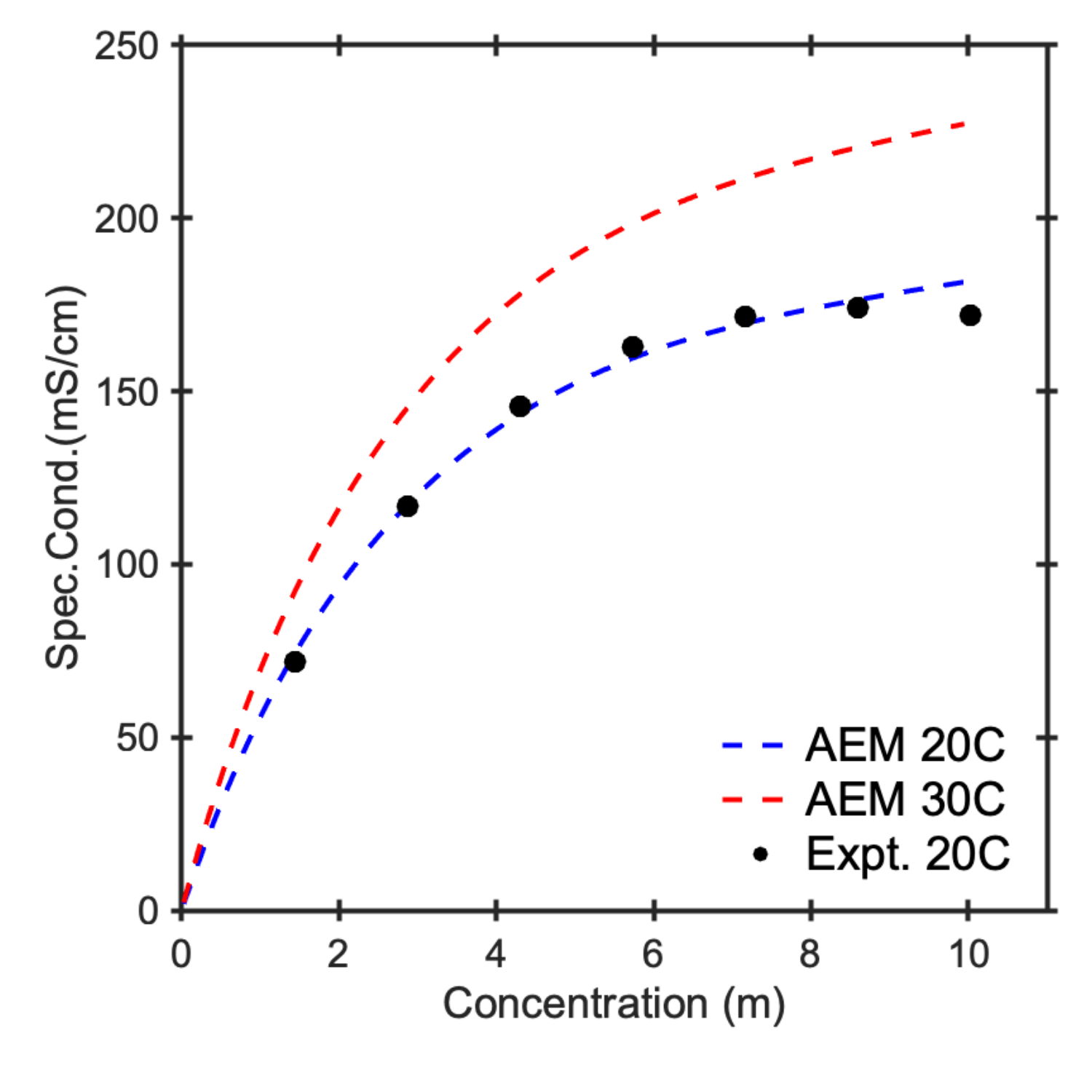}
        \caption{\ce{NaNO3}}
    \end{subfigure}
    ~ 
    \begin{subfigure}[b]{0.4\textwidth}
        \includegraphics[width=\textwidth]{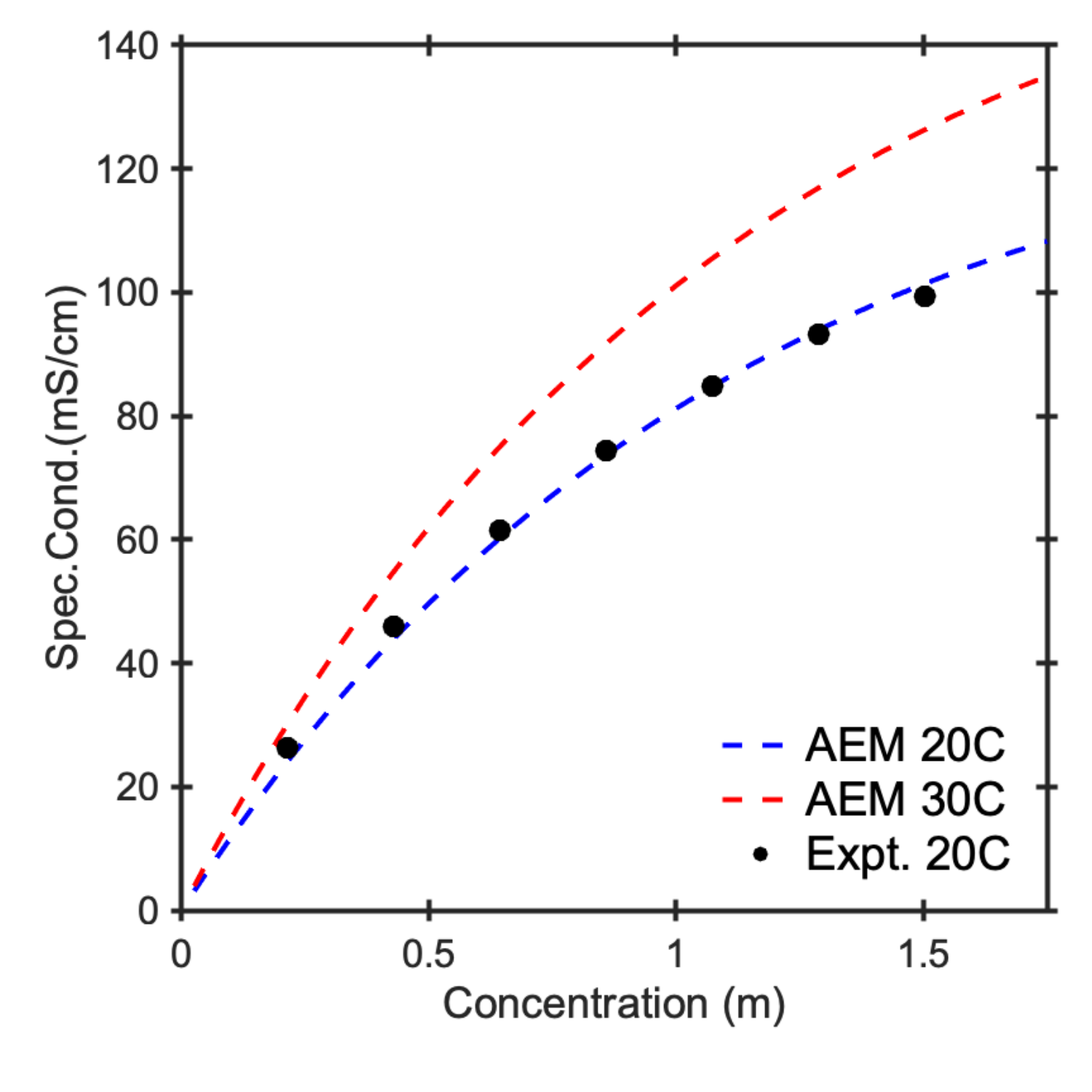}
        \caption{\ce{Na2SO4}}
    \end{subfigure}
    
    \begin{subfigure}[b]{0.4\textwidth}
        \includegraphics[width=\textwidth]{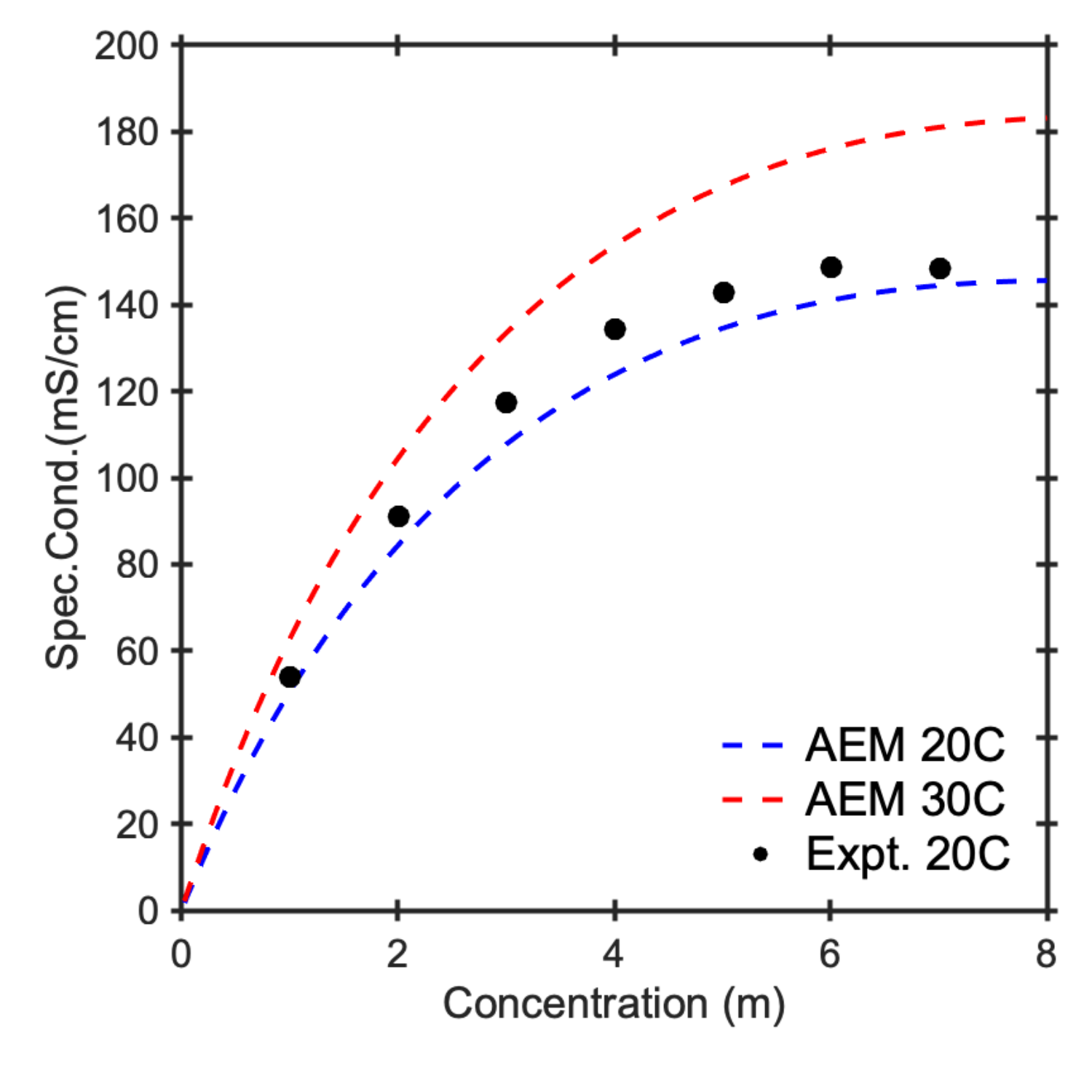}
        \caption{\ce{LiNO3}}
    \end{subfigure}
    \begin{subfigure}[b]{0.4\textwidth}
        \includegraphics[width=\textwidth]{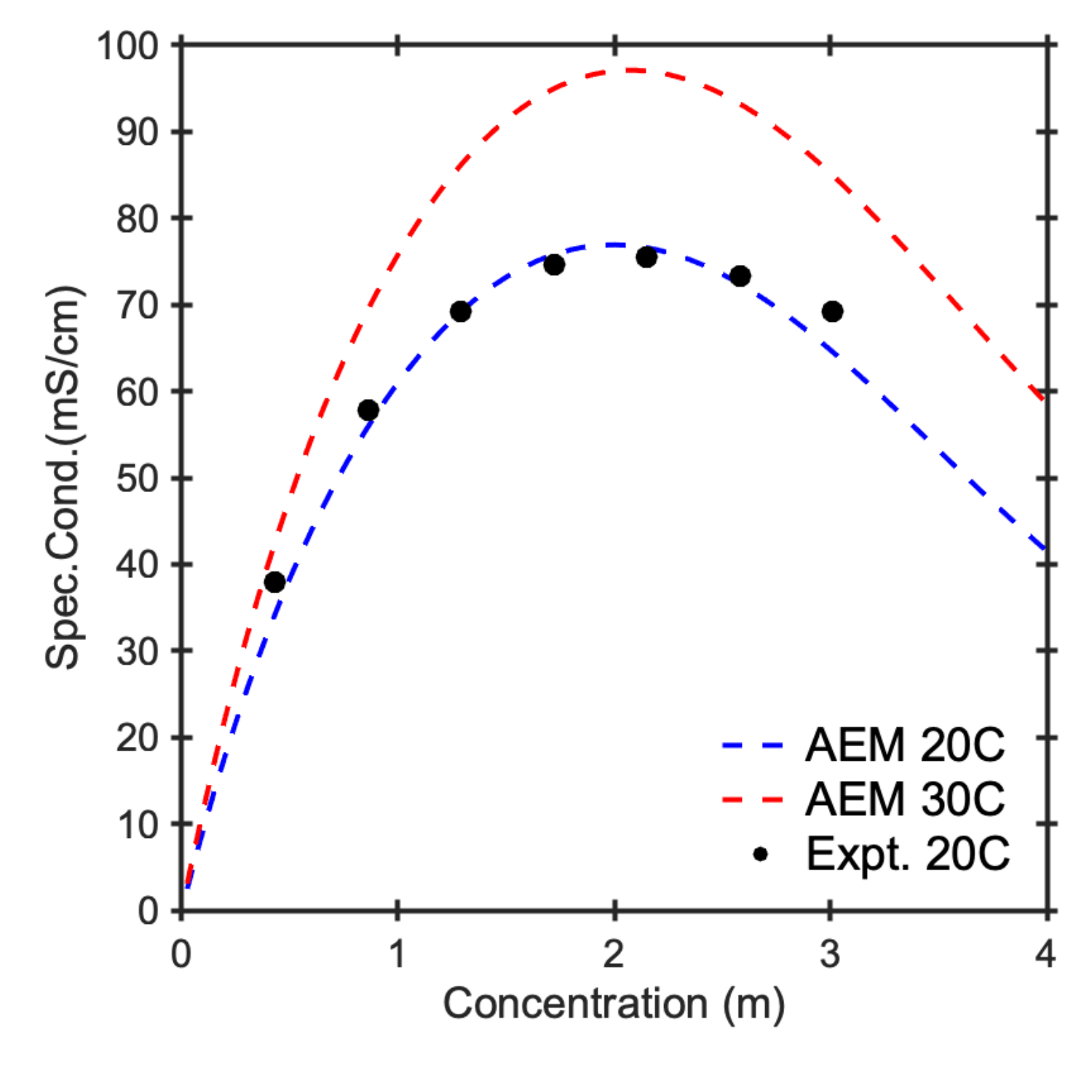}
        \caption{\ce{Li2SO4}}
    \end{subfigure}

    \caption{Measured (black) vs AEM-calculated (red/blue) conductivities for four salts.  The predictions of AEM are accurate to within 5\% on average when compared to experiments.}
\end{figure}

\begin{figure}
    \centering
    \begin{subfigure}[b]{0.5\textwidth}
        \includegraphics[width=\textwidth]{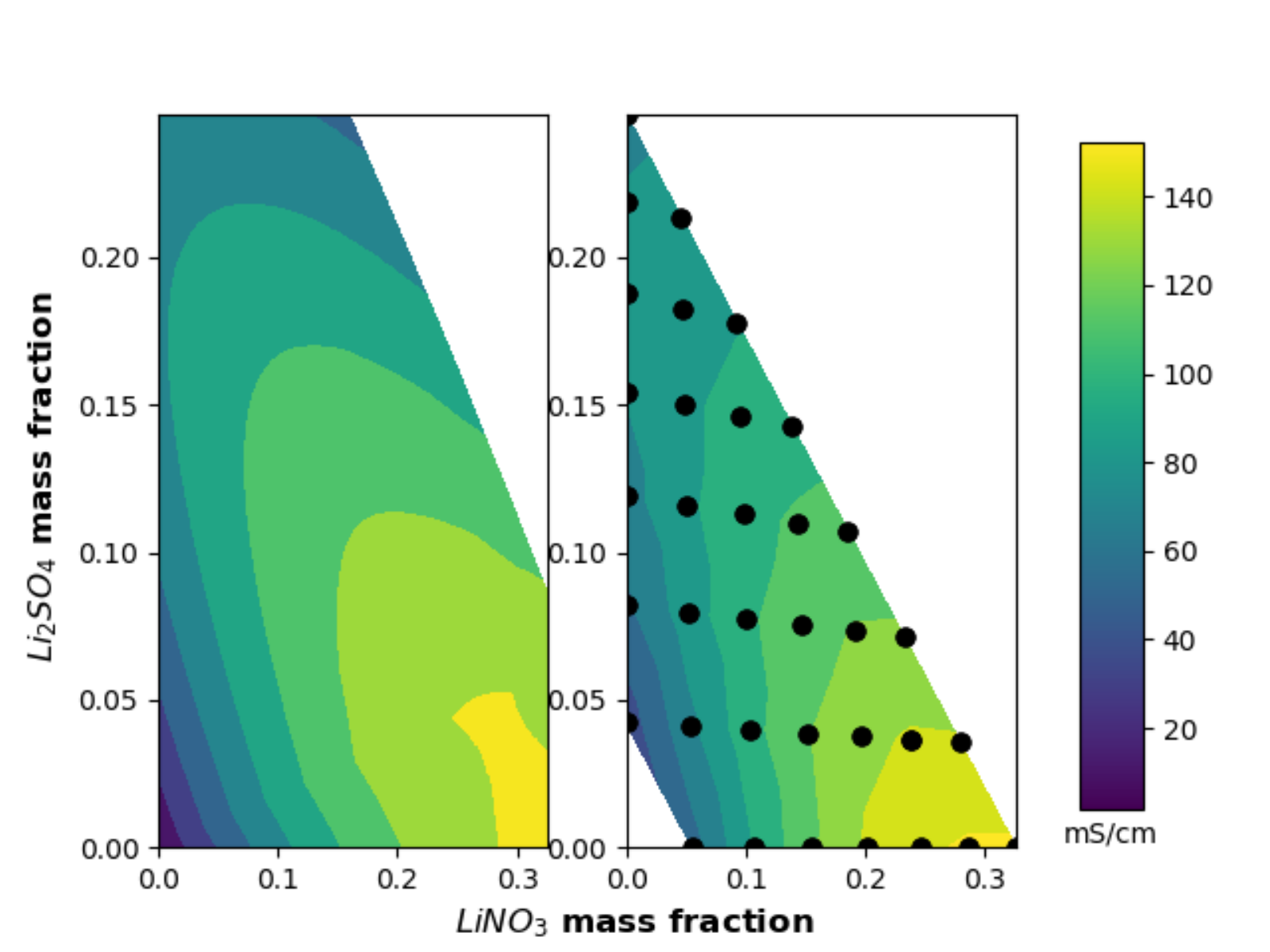}
        \caption{\ce{LiNO3}/\ce{Li2SO4}}
    \end{subfigure}
    
    \begin{subfigure}[b]{0.5\textwidth}
        \includegraphics[width=\textwidth]{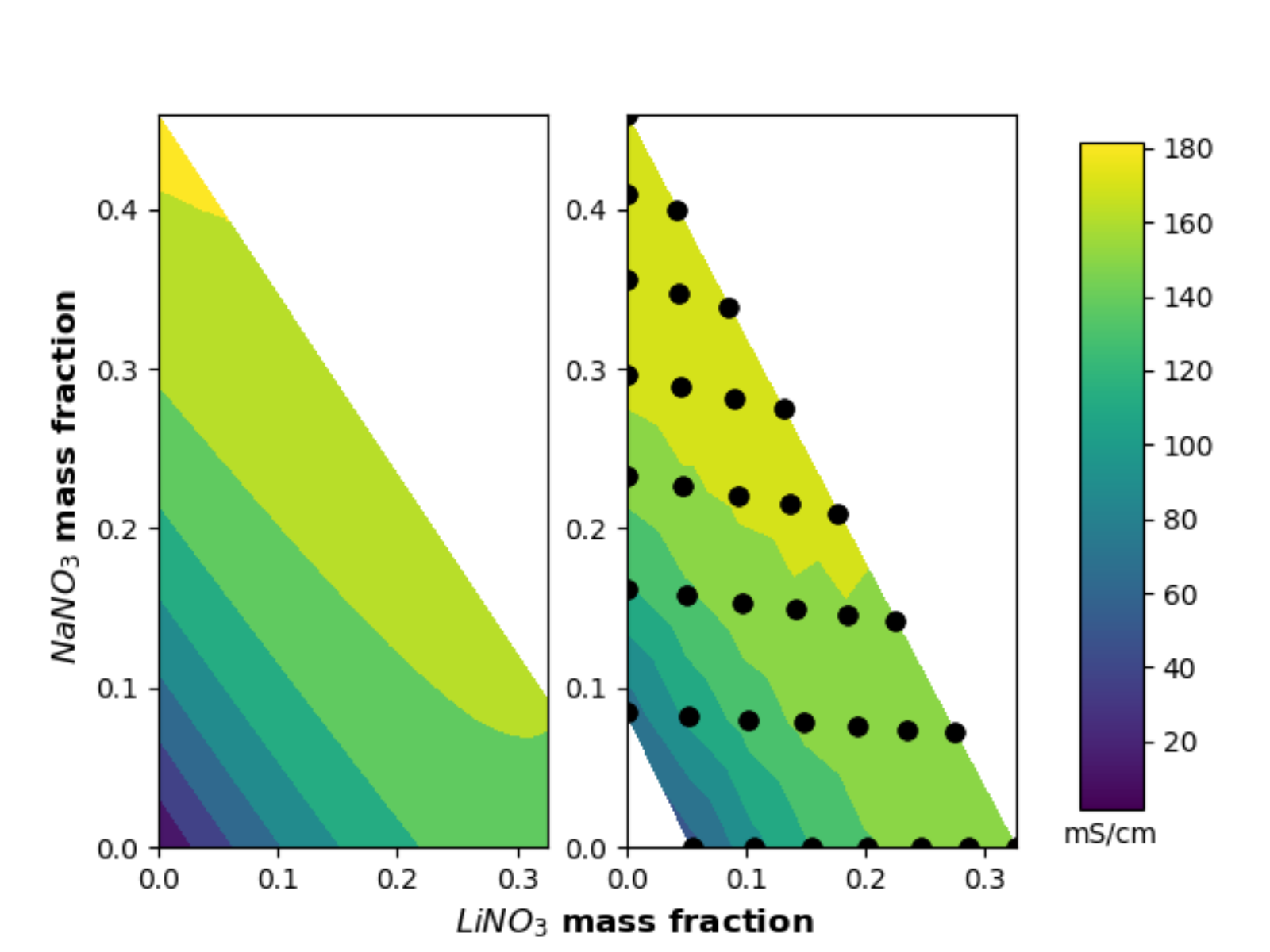}
        \caption{\ce{LiNO3}/\ce{NaNO3}}
    \end{subfigure}
    
    \begin{subfigure}[b]{0.5\textwidth}
        \includegraphics[width=\textwidth]{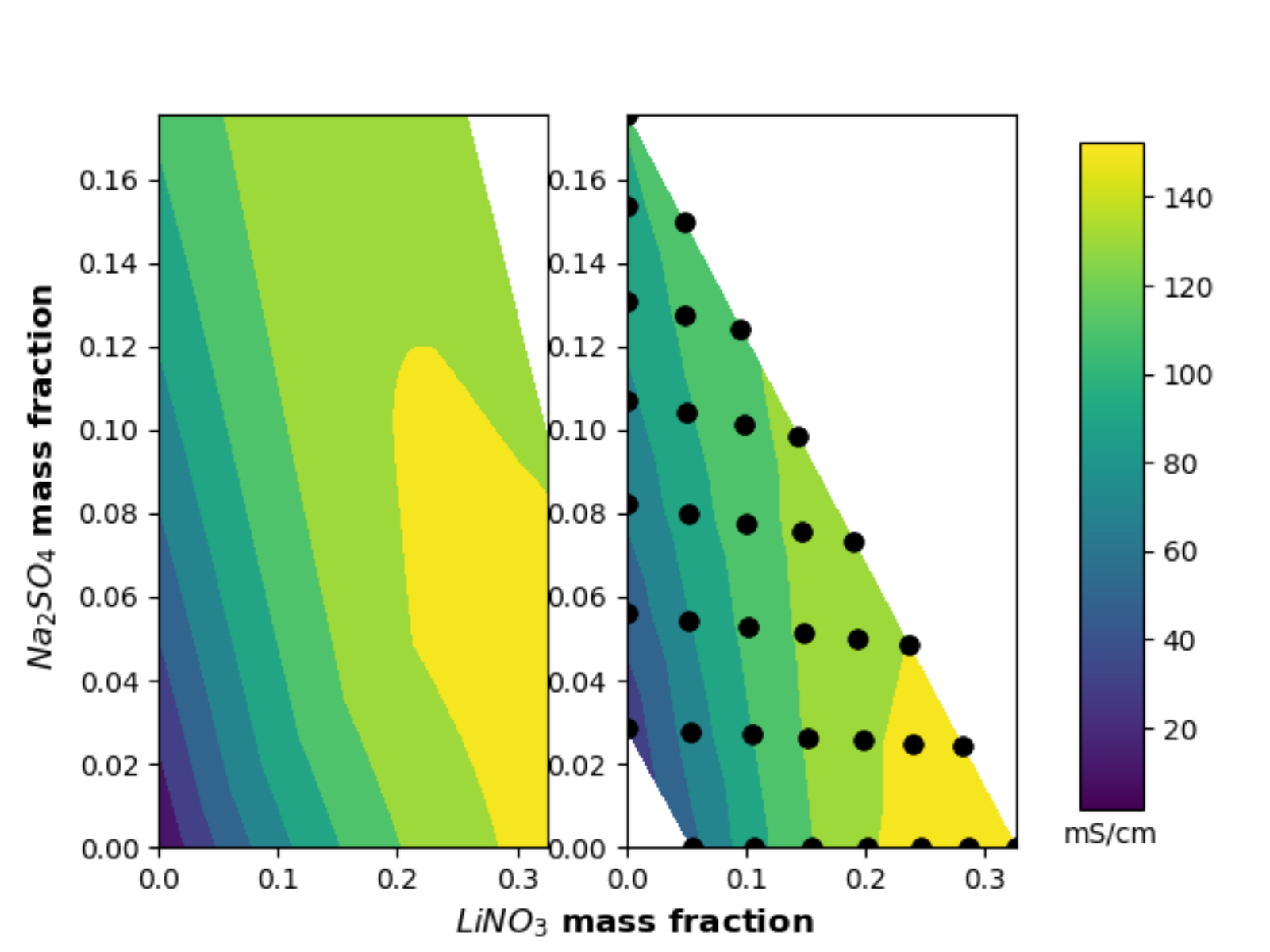}
        \caption{\ce{LiNO3}/\ce{Na2SO4}}
    \end{subfigure}
    \caption{Measured (right, black dots for measured points) vs AEM-calculated (left) conductivity for three binary salt mixtures.  The AEM predictions are within 10\% in the entire phase space when compared to experiments.}
\end{figure}

Figures 1 (a)-(d) compare bulk specific conductivities from AEM against experiment for \ce{NaNO3}, \ce{Na2SO4}, \ce{LiNO3}, and \ce{Li2SO4}. AEM data at both 20\textdegree C and 30\textdegree C are plotted against test-stand measurements taken between 21\textdegree C and 23\textdegree C, compensated by the Consort meter to 20\textdegree C.

AEM calculations of \ce{NaNO3} and \ce{Na2SO4} conductivities show very close agreement with experiment. The most significant divergence (approximately 9 mS/cm over-estimation of conductivity) appears at the highest molalities of \ce{NaNO3}. Slight divergence is shown near the lowest and highest molality measurement in \ce{Na2SO4}, but this is merely 2 mS/cm.

AEM calculations of \ce{Li2SO4} also show very close agreement with experiment. Like in the case of \ce{NaNO3}, the most significant divergence (approximately 5 mS/cm under-estimation of conductivity) appears at the highest molalities of \ce{Li2SO4}. Elsewhere, divergence is less than 2 mS/cm.

AEM calculations of \ce{LiNO3} are in close agreement with experiment. The most significant divergence is close to 10 mS/cm near 4 mol/kg water. However, this error of less than 10\% the expected value is less than the most significant errors shown in concentrated non-aqueous system validations \cite{Gering2017}.

Qualitatively, AEM captures the behavior of sulfate electrolyte conductivity more accurately than nitrate electrolytes. In the case of the sulfates, AEM accurately predicts the maximum of conductivity on the concentration curve. This is not the case in nitrates, where AEM predicts the maximum of conductivity past the concentration which experiments determined to be the solubility limit. Diminishing conductivity in high-concentration electrolytes may come from increased viscosity of the liquid. Thus, the divergence in prediction in the \ce{Li2SO4} and \ce{NaNO3} at high concentrations may be due to viscosity effects that AEM does not precisely capture in this regime. While AEM has shown great accuracy in viscosity estimation in chloride aqueous electrolytes, up to molarities of 8 in \ce{LiCl} systems\cite{Gering2006}, viscosity has not been benchmarked for the anions used in this study. Further evaluation of viscosity predictions from AEM in a broad range of aqueous electrolytes would be required to test this hypothesis.

Figures 2 (a) - (c) compare bulk specific conductivities from AEM against experiment for binary mixtures of \ce{LiNO3} with \ce{Li2SO4}, \ce{LiNO3} with \ce{NaNO3}, and \ce{LiNO3} with \ce{Na2SO4}. The colored contour lines show conductivity calculated/measured for an electrolyte composition, as described by mass fraction of the electrolyte composed of each salt species; black dots on the experimental data figures (right side) correspond to electrolyte compositions experimentally evaluated. 

AEM calculations of binary mixtures of \ce{LiNO3} and \ce{Li2SO4} give very strong qualitative descriptions of the system conductivity, and strong quantitative agreement compared to experimental results. The general behavior of the mixed conductivity system is very accurately described by AEM. AEM predicts the maximal conductivity being found in highest-concentration \ce{LiNO3}, which is confirmed by experimental data. AEM accurately predicts compositions along the solubility limit to be higher in conductivity than less dilute systems. Generally, the AEM-calculated shape of conductivity behavior in highly mixed systems is replicated in experiments. Quantitatively, the accuracy of the model drops in the high-concentration, mixed cases (e.g. lower right quadrant of experimental results) to an error of approximately 10 mS/cm.

AEM calculations of binary mixtures of \ce{LiNO3} and \ce{NaNO3} give very strong qualitative results and good quantitative results compared to experimental results. AEM agrees with experiment on the system's maximal conductivity being the highest-concentration \ce{NaNO3}. AEM also accurately predicts compositions along the solubility limit to be higher in conductivity than less dilute systems. The AEM-calculated shape of conductivity behavior is again replicated by experiments across the composition space. A key area of quantitative disagreement is in the mixed, high-concentration \ce{LiNO3} region near the solubility limit, where errors are approximately 10 mS/cm.

AEM calculations of binary mixtures of \ce{LiNO3} and \ce{Na2SO4} give strong qualitative results and good quantitative results compared to experimental results. AEM predicts the system's maximal conductivity to be in highest-concentration \ce{LiNO3} space, which is supported by experimental data. AEM also accurately describes compositions with more \ce{LiNO3} to always be higher in conductivity than systems with less (i.e. generally moving rightward in the composition space). Key areas of disagreement include the space with high \ce{LiNO3} mixed with \ce{Na2SO4}, where errors again are approximately 10 mS/cm.

Our results indicate that, while the mixing parameters set in AEM could potentially be improved in certain cases (e.g. \ce{LiNO3} with \ce{Na2SO4}), the chemical physics framework for describing these systems is highly accurate even in highly-concentrated, mixed systems. Most importantly, AEM correctly identifies the behavior and trend of mixed systems, with quantitative results within 10\% of true values across the board. These errors are lower than previous conductivity benchmarks of AEM in non-aqueous systems.\cite{Gering2017}  It is very important to highlight that this method is more accurate when compared to both quantitative and qualitative errors from predicting conductivity using computing-intensive atomistic modeling (e.g. molecular dynamics using classical or polarizable force fields)\cite{Khetan2018,Picalek2007,Lee2005}.

Our work suggests that AEM can be a useful tool for screening potential aqueous electrolyte formulations. Due to its robust prediction of transport properties in highly-concentrated, mixed aqueous electrolytes, large fractions of the electrolyte design space can be filtered without experimental testing. Fewer experimental evaluations can lead to development of high-performing aqueous electrolytes in less time and less development cost. Future work will seek to automate this electrolyte design process, incorporating AEM predictions as a prior in a Bayesian optimization framework coupled to the robotic test-stand.

\section{Conclusion}

Promising candidates for an aqueous electrolyte with a wide voltage stability window feature high concentrations of lithium or sodium salts. An effective screening tool for these candidates would accurately predict transport properties for these aqueous electrolyte systems. In this work, we have measured conductivity of unary and binary aqueous electrolyte systems and the experimental data indicates that AEM accurately predicts conductivity up to high-concentration for four single salt systems of interest to aqueous electrolytes. We have also shown accurate prediction of conductivity in three binary salt systems. Maximum errors for AEM against experiment were near 10\%, smaller than some maximum errors in previous conductivity benchmarks of AEM in concentrated non-aqueous systems \cite{Gering2017}. Maximum deviations tended to occur in highly-concentrated and highly-mixed regimes, indicating potential improvements that could be made in the viscosity models or mixing parameters for AEM. 

Our results indicate that AEM's chemical physics framework provides robust predictions of conductivity in complex, non-ideal aqueous systems. Compared to atomistic methods, AEM provides much higher predictive accuracy and significant savings in computing cost. Further work will utilize AEM as a screening tool in the automated design of high-performance aqueous systems, intending to yield safer and lower-cost battery systems.

\begin{acknowledgement}
This work was supported by Toyota Research Institute through the Accelerated Materials Design and Discovery program.
\end{acknowledgement}

\clearpage

\bibliographystyle{achemso} 
\bibliography{aem_benchmark} 

\end{document}